\documentclass[letterpaper]{ptephy}

\begin{document}

\newcommand {\beq}{\begin{eqnarray}}
\newcommand {\eeq}{\end{eqnarray}}
\newcommand{\tc}{\textcolor}
\newcommand{\blue}{\textcolor{blue}}


\title{Radion Inflation in Higher-Dimensional Gravity Theory}

\author{\name{Yuichi Fukazawa}{1,\ast}, \name{Takeo Inami}{1}, and
\name{Yoji Koyama}{1}}

\address{\affil{1}{Department of Physics, Chuo University, Bunkyo-ku, Tokyo 112, Japan}
\email{fukazawa@phys.chuo-u.ac.jp}}

\begin{abstract}
We study a radion model for inflation and density perturbation identifying the scalar field arising from 
the five-dimensional gravity on $M_4 \times  S^1$ with the inflaton.
 The inflaton potential arising from 5D graviton and matter one-loop effects is
  shown to possess the desirable slow-roll property. This radion model
   contains only a small number of parameters and there is an ample parameter region 
   which reproduces (without fine-tuning) all precision astrophysical data. 
\end{abstract}

\subjectindex{E81,E84}
\maketitle
\section{Introduction}
The Higgs field(s) and the inflaton play significant roles in particle physics and cosmology.
 As to their origins there are a few different views.
  One attractive view is that they are both extra-dimensional components of gauge fields in higher dimensions \cite{1,2}. 
  Some time ago we have constructed an inflaton model in this picture and studied their astrophysical consequences \cite{3,4}.
   We immediately note that graviton in higher dimensions also contains scalar fields if extra dimensions are compactified \cite{H}. 
   It is then an interesting possibility that the inflaton is a scalar field of this type. 
   
In this paper we pursue this possibility studying the radion model of inflaton identifying 
     the scalar field arising from the five-dimensional (5D) gravity on $M_4 \times  S^1$  
     ($M_4$ is the 4D Minkowski spacetime) \cite{5} with the inflaton. 
     The inflaton potential then arises from 5D graviton and matter one-loop effects.
      The question then is whether the parameter values of the model have a region 
      in which the inflaton potential possess the property 
      in accord with the precision astrophysical data (hopefully without fine-tuning).
      
 A few people have previously studied the possibility of 
the radion giving rise to cosmological inflation \cite{K,T}. Our 
model is different from the previous ones in the way that 
the inflaton potential is derived from a particle physics model; the potential is 
generated from the gravity and matter quantum effects. 
Hence it is fixed from 5D gravity theory, modulo two 
parameters of this theory. 

It is possible that the scalar field from gauge field and that from the graviton may compete in explaining the inflation.This work is a first step to address this question.
     
The potential of the scalar fields $\phi_i$ from the extra dimensions of the graviton is due to the graviton loop. 
     It is known to be unbounded from below toward $\phi_i=0$ \cite{5}.
      This instability can be cured by considering fermion matter loops \cite{6,7}.
       We consider 5D gravity plus matter on $M_4 \times  S^1$, and use the one-loop potential evaluated in \cite{5}. 
       Our model contains only a small number of parameters, the $S^1$ circumference $L$, the fermion mass $\mu$ and the number of fermions $c$. 
       
The vacuum energy is positive, so-called dark energy (positive cosmological constant) in the inflation period. 
     It means that the space-time is de-Sitter like. We are not concerned with this issue and deal with loop corrections in 4D Minkowski space-time.
\section{The Model and One-Loop Effective Potential}
We consider the five-dimensional (5D) gravity theory with matter fermions  $\psi_i$ compactified on a circle $ S^1$ of circumference $L$.
Here $i=1,\cdots ,c$, $c$ being the number of fermions.
We denote the 5D coordinate by $ x^M = (x^\mu , y) $, 
the index $M$ running from 0 to 3, 5, and the index $\mu$ from 0 to 3.
The action in 5D is given by 
\begin{equation}
S=\int  d^5 x \sqrt{g} \left [\frac{1}{16 \pi G_5}  R_5+ \bar{\psi_i} \left ( \mathit{i} \gamma^{A} D_{A} - \mu  \right) \psi_i \right] ,
\end{equation}
\begin{equation}
g_{MN} = \Phi ^{-1/3} \left(
\begin{array}{cc}
g_{\mu \nu}+A_\mu A_\nu  & A_\mu \Phi  \\
A_\nu \Phi & \Phi  
\end{array}
\right) ,
\end{equation}
where $R_5$ is the 5D scalar curvature, $G_5$ is the 5D gravitational coupling constant.
5D fermions $\psi_i$ have the same mass $ \mu $.
$ g_{MN}$ contains a scalar field $\Phi$ after compactification on $S^1$.
The zero mode of $\Phi$, $\Phi^{(0)} \equiv \phi$, is a 4D scalar field called the radion. Note that $\phi$ is dimensionless because it is a part of the metric $g_{MN}$.
We identify the radion $\phi$ with the inflaton. 
Thus, we do not introduce extra scalar fields in order to construct the inflation model.

The physical circumference $L_{\rm phys}$ of the fifth dimension is given by 
\begin{equation}
 L_{\rm phys}= \int^L _0 dx^5 \sqrt{g_{55}} = L\phi^{1/3},
\end{equation}
where $L=2 \pi \times (S^1   {\rm radius} )$,
 meaning that $L_{\rm phys}$ is determined by $\langle \phi \rangle$, the VEV of the $\phi$ and the adjustable parameter $L$. 
Hence the dynamics of the extra dimension is described by the effective potential $V(\phi)$ of the radion $\phi$.
We evaluate $V(\phi)$  to study the behavior of the fifth dimension.

The potential $V(\phi)$ is generated by the radiative corrections of 5D graviton and matter one-loop. 
To this end, we introduce the classical and quantum fields.
\begin{equation}
g_{\mu \nu}= \delta_{\mu \nu} + h_{\mu \nu}, \quad 
\phi=\phi_c+\phi_q , \quad 
A_\mu = 0+A_\mu  ,
\end{equation}
where $ \delta_{\mu \nu} $, $ \phi_c $ are the classical fields and $ h_{\mu \nu} $, $ \phi_q $ and $ A_\mu $ are the quantum fields.
The effective potential for the radion $\phi$ at the one-loop level is obtained in \cite{5,6}.
\begin{eqnarray}
 V(\phi)&=&V_{\rm grav}(\phi) + V_{\rm mat}(\phi)\\ \nonumber
      &=& - \frac{3\zeta(5)}{64\pi^2} \frac{1}{L^4 \phi^2}+c \frac{3}{64 \pi^2} \frac{1}{L^4 \phi^2}
       \left[ Li_5 \left(e^{-2L \mu {\phi^{1/3}}} \right) 
     + 2L\mu\phi^{1/3} Li_4 \left(e^{-2L \mu \phi^{1/3}} \right) \right. \\ 
     &&  + \left. \frac{4}{3} L^2 \mu^2 \phi^{2/3}  Li_3 \left(e^{-2L \mu \phi^{1/3}} \right) \right] 
       +aL\phi^{-1/3} + b .\label{pot}
\end{eqnarray}
The first term in the second equality is the contribution from the 5D gravity loop.
The second term is that from the matter loop, with $c$ being the number of fermions.
In the third line, $ aL\Phi^{-1/3} $ is the 5D cosmological constant counter term and $b$ is a constant term which we will explain later.
$Li_n$ ($n$ is an integer) is the polylogarithm function defined by   
\begin{equation}
Li_n(u)= \sum_{k=1}^{ \infty } \frac{u^k}{k^n}.
\end{equation}

It is known that the one-loop effective potential in the pure gravity sector is attractive 
and that $V_{\rm grav}(\phi)$ tends to negative infinity as $\phi \rightarrow 0$,
indicating that the perturbative computation cannot be trusted near $\phi=0$. 
Rubin and Roth pointed out that the stable minimum of the potential is achieved by introducing matter fermions \cite{7}.
\begin{figure}[t]
\begin{center}
\begin{minipage}{12cm}
\begin{center}
\includegraphics[height=4.5cm,width=8.7cm,angle=0,clip]{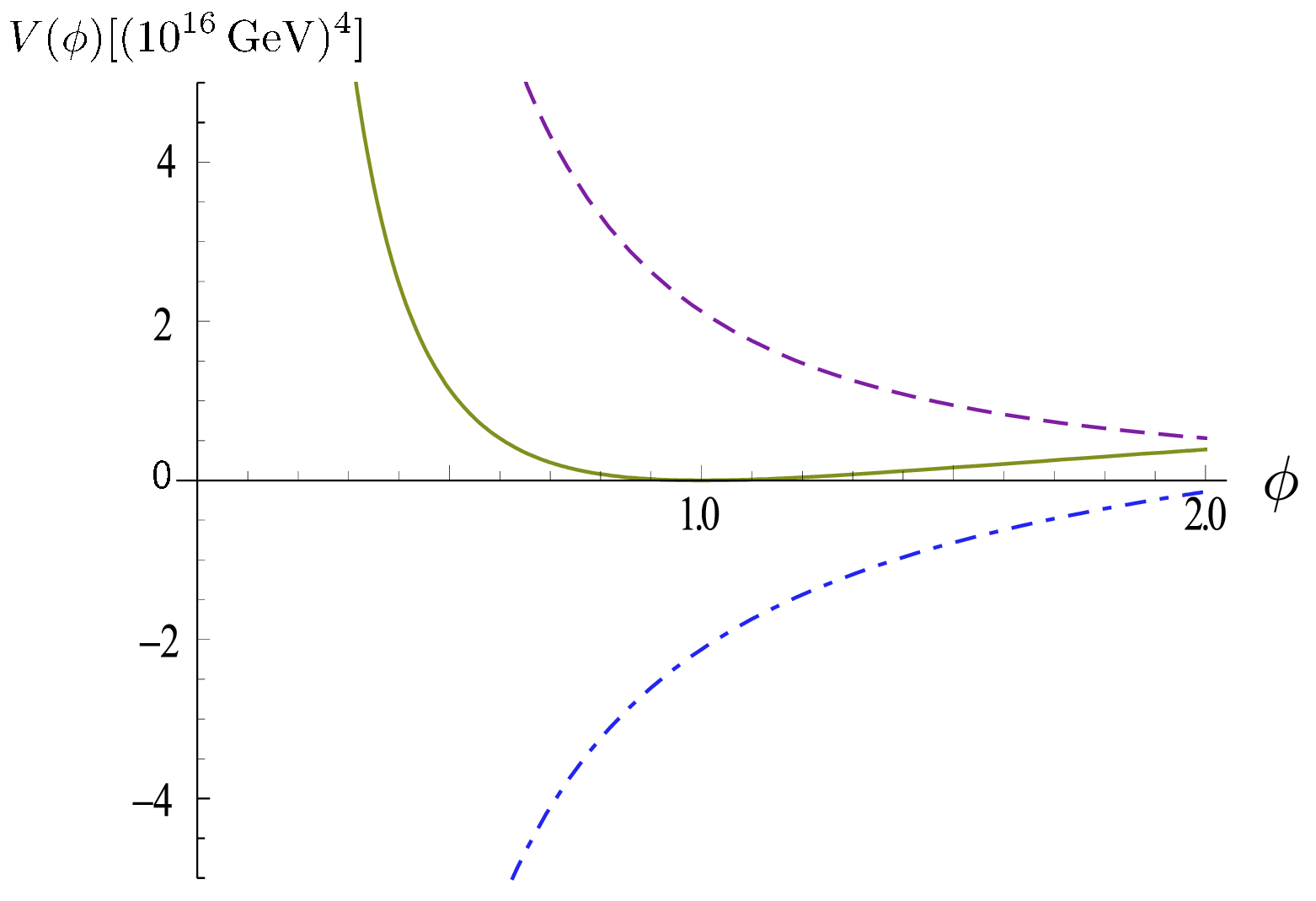}
\caption{
The effective potential arising from the graviton 1-loop (double broken line) and the matter 1-loop (broken line) for $c=2$.
The sum of the two has a stable minimum (solid line).}
\label{fig:}
\end{center}
\end{minipage}
\end{center}
\end{figure} 
We compare the attractive potential $V_{\rm grav}$ arising from graviton 1-loop with the repulsive potential $V_{\rm mat}$ from the matter 1-loop in fig.1.
$V_{\rm mat}$ shown in fig.1 is the case of two matter fermions ($c$=2).
The two terms compete with each other.
The sum of the two terms gives the potential $V(\phi)$ which is unbounded or bounded from below depending on the number of fermions $c$. 
$V(\phi)$ at small $\phi$ can be found by noting
$ Li_n \left(e^{-2L \mu {\phi^{1/3}}} \right) \simeq \zeta(n)-2L \mu \phi^{1/3} \zeta (n-1)$.
We have 
\begin{eqnarray}
V(\phi) \simeq - \frac{3\zeta(5)}{64\pi^2} \frac{1}{L^4 \phi^2}+c \frac{3\zeta(5)}{64\pi^2} \frac{1}{L^4 \phi^2}.
\end{eqnarray}
On the other hand $V_{\rm grav}(\phi)$ dominates the potential for large values of $\phi$,
\begin{eqnarray}
V(\phi) \simeq  - \frac{3\zeta(5)}{64\pi^2} \frac{1}{L^4 \phi^2}+aL\phi^{-1/3}.
\end{eqnarray}
Hence the potential (\ref{pot}) can acquire a minimum at some finite $\phi$ ($\phi=1$ in fig.1) by choosing $ \ c \geq  2$ (two or more fermions).
\section{Radion inflation in five dimensional gravity with matter theory}
 In this section, we investigate the constraints on our inflation model from the observational data and the theoretical consideration. 
We will see whether our proposal of the inflaton and the radion being the same field 
has a meaning in comparison with recent astrophysics observation.
  \subsection{Constraints on the inflation model}
 We recapitulate the conditions which the inflation model should satisfy \cite{8,9}. 
\begin{enumerate}
\item[i)] The slow-roll condition
  \begin{equation}
  \epsilon \equiv \frac{1}{2}M^2_P \left(\frac{V'}{V} \right)^2 \ll 1      \: , \: \quad
 \eta \equiv M^2_P \frac{V^{''}}{V} \ll 1 ,
  \end{equation}
 where $M_P= 2.44\times 10^{18} \mathrm{GeV} $ is the reduced Plank mass, and $V'=dV/d\phi$.
\item [ii)]The spectral index $n_s$ is known quite precisely \cite{9}.
\begin{equation}
n_s \equiv 1-6\epsilon +2\eta  \: , \:  \quad  0.948<n_s<0.977 .
\end{equation}
\item[iii)] The number of e-foldings $N$ 
\begin{equation}
N \equiv \frac{1}{M^2_P} \left| \int ^{\phi_e} _{\phi_i} \left(\frac{V}{V'} \right) d \phi \right| = 50-60,
\end{equation}
where $\phi_i$ is  the value of $\phi$ at the beginnnig of the inflation and $\phi_e$ at its end.
For solving the flatness and the horizon problems the inflation must last for a sufficiently long time,
namely $N$ has to be $50-60$.
 \item[iv)] The curvature perturbation \cite{9}
\begin{equation}
\delta _H = \frac{1}{5\sqrt{6}\pi} \frac{V^{1/2}}{M^2_P  \epsilon^{1/2}} = 1.91 \times 10^{-5}.
\end{equation}
\item[v)] There is an upper bound for the tensor-to-scalar ratio $r$ from the observations \cite{9}.
\beq
r=16\epsilon \lesssim 0.24.
\eeq
 \item[vi)] The vacuum energy is nearly zero, $V_{\rm min}=0$.
The value of the $V_{\rm min}$ corresponds to the cosmological constant and it is   known to be positive and nearly zero. 
\end{enumerate}

We study whether the conditions for inflation summarized above are met for reasonable values of the parameters of our model, $L$, $\mu$ and $c$.
  \subsection{Numerical analysis of our radion inflation model}
 We note that there is only one minimum in $V\left(\phi \right)$, $\phi=\phi_c$ (see $V(\phi)$ in fig.1).
We consider the case in which the radion field $\phi_f$ is fluctuating around $\phi_c$, and set
\begin{equation}
\phi = \phi_c +  \frac{\phi_f}{\sqrt{3} M_{P}}. 
\end{equation}
The physical size of the fifth dimension is $L_{\rm phys}=\phi_c^{1/3} L$. Note that $\phi_c$ is dimensionless and $\phi_f$ has the canonical mass dimension. 

We suppose that our model is  a low energy effective theory below the Planck scale.
The result that the fifth dimension for our radion-inflaton model is stabilized near the Planck scale (see fig.1) implies that the model is a consistent effective theory. In this model, three spatial dimensions expand during the inflation whereas the fifth dimension shrinks to the Planck scale.

We now address ourselves to a more quantitative question in comparison with the constraints explained in subsec.\,3.1. The question is whether there is such parameter ($L,\, \mu,\, c$) values that the astrophysical data are reproduced. To this end we fit the data by varying the parameters. The numerical analysis is not so straightforward as it first looks. This is because the potential depends on five parameters even after choosing $c=2$, $V=V(\phi_f;\, \phi_c,\, L,\, \mu,\, a,\, b)$. We have handled this complicated situation by first fixing the minimum of $V$ to eliminate two parameters $a$ and $b$. We impose two conditions
\beq
   V(\phi_f;\, \phi_c=1,\, L,\, \mu,\, a,\, b)=0,\quad
   V'(\phi_f;\, \phi_c=1,\, L,\, \mu,\, a,\, b)=0. 
   \eeq
The first condition means the zero vacuum energy and the second that $V$ has the minimum at $\phi_c=1$. These two conditions are used to eliminate $a$ and $b$ in favor of $L$ and $\mu$, 
\beq
   a=a(L,\, \mu),\ \ b=b(L,\, \mu).
\eeq    
The potential is expressed as $V(\phi_f;\, L,\, \mu)$. We then look for values of $L$ and $\mu$ for which the conditions i) $-$ v) are obeyed. 

We first choose the value of $L$. We envisage the situation in which $L$ is of the order of $1/M_P
\, (=4.18 \times 10^{-17} {\rm GeV^{-1}})$. The $c$, the number of fermions $\psi_i$, may be chosen as we like. However, a huge number of $\psi_i$ seems unrealistic. Here we tentatively take $c=2$, the smallest value which is consistent with the stability (bounded from below) of the potential. As for $\mu$, we have a priori little idea about it's magnitude.
\begin{figure}[t]
 \begin{minipage}{0.49\columnwidth}
  \begin{center}
   \includegraphics[height=49mm,width=74mm]{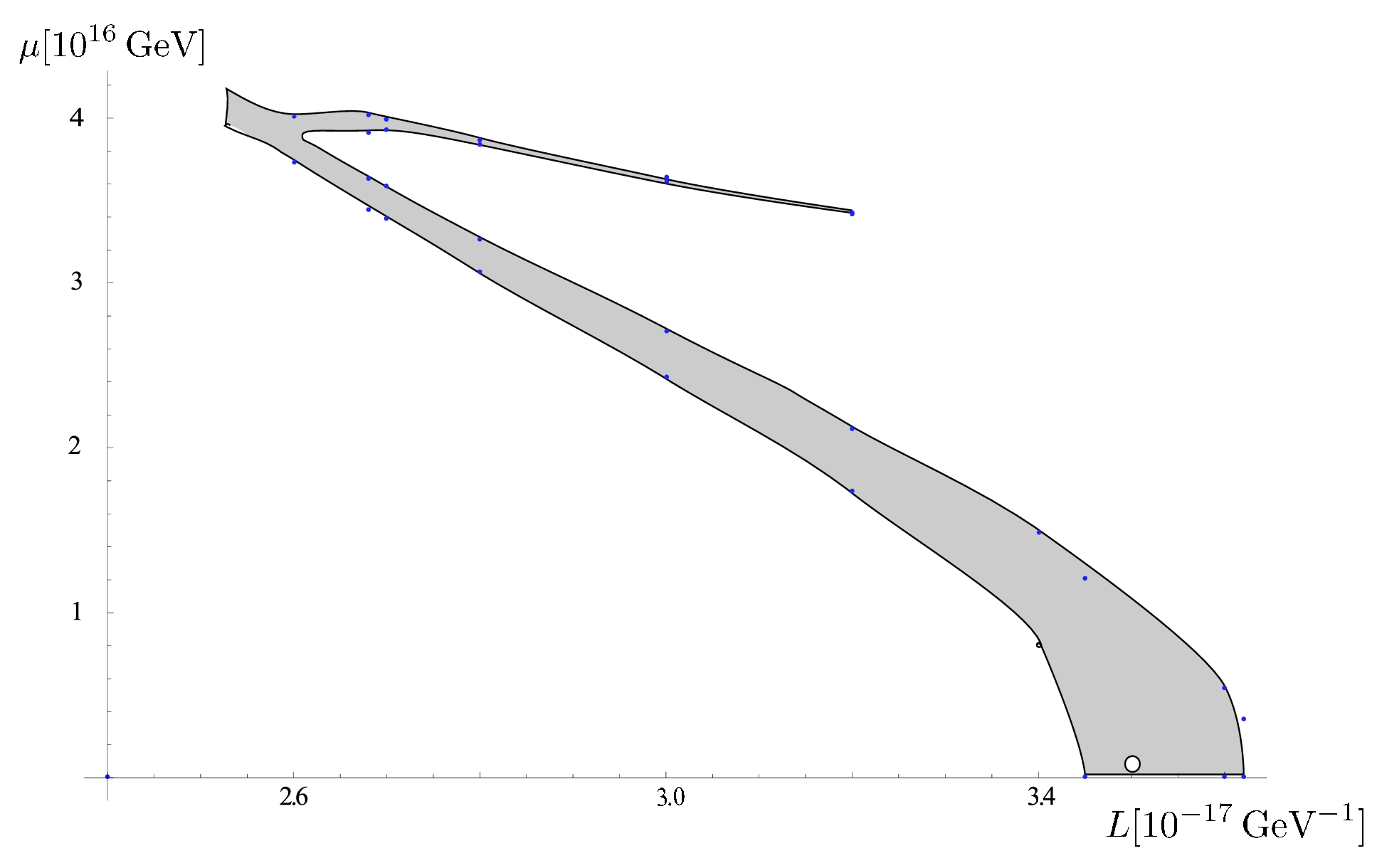}
  \end{center}
  \caption{The allowed region of the parameters is shown by a shaded area. We can choose the $L$ and the matter mass $\mu$ from inside of the shaded area, $L=(2.53-3.62)\times 10^{-17}  \mathrm{GeV^{-1}}$ and $\mu=0.8-4.15\times 10^{16}  \mathrm{GeV}$.}
  \label{fig:one}
 \end{minipage}
 \hspace{0.18cm}
 \begin{minipage}{0.5\columnwidth}
  \begin{center}
   \includegraphics[height=43mm,width=76mm]{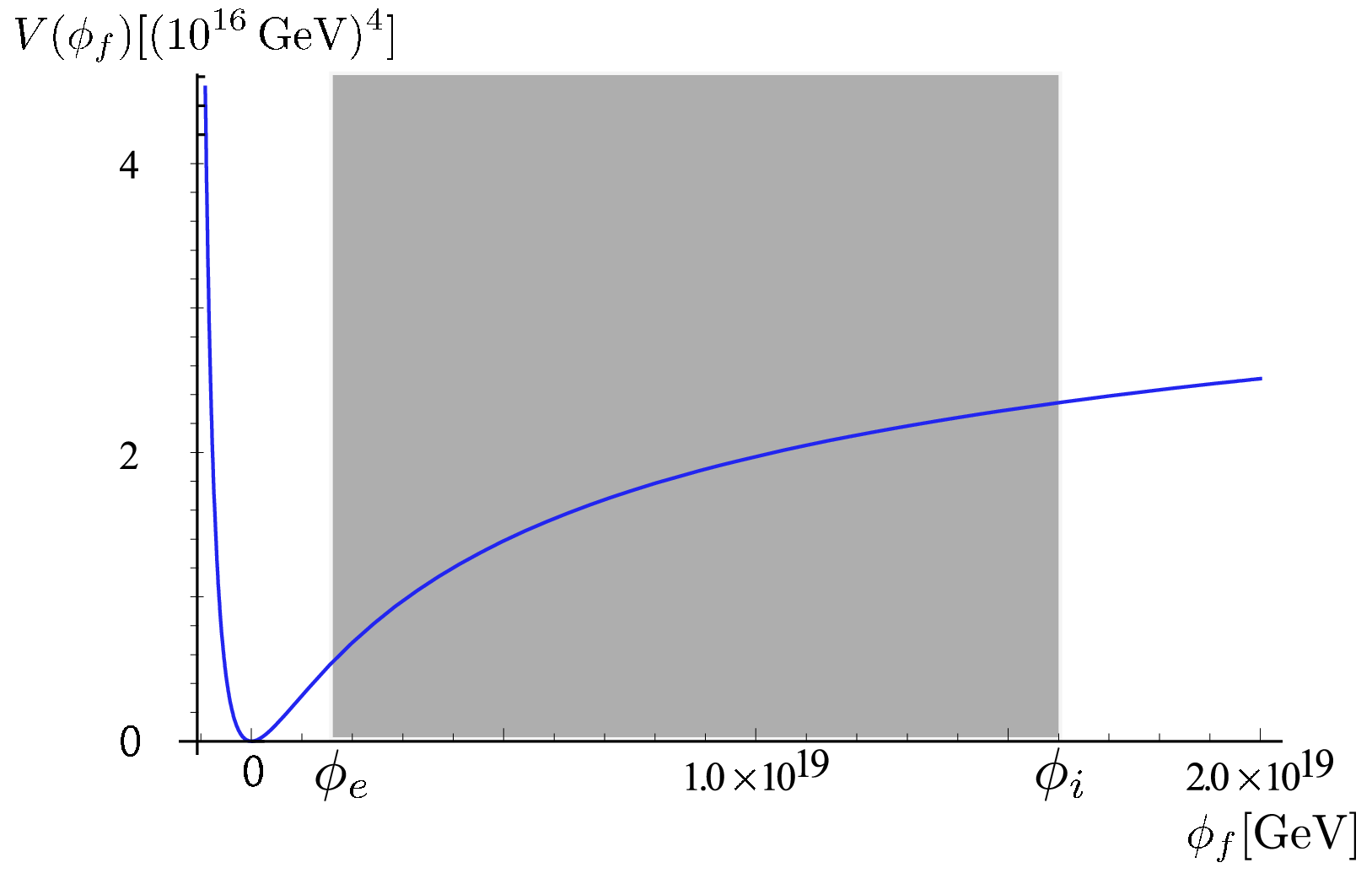}
  \end{center}
  \caption{The inflaton potential and the inflation period for typical values of $L$ and $\mu$. The values (\ref{ex}) of $L$ and $\mu$ are used. }
  \label{fig:two}
 \end{minipage}
\end{figure}

{\bf I)} Taking $c=2$, there remain only two parameters, $L$ and $\mu$. We try to fit all the astrophysical data, i) $-$ v) of subsec.\,3.1, allowing both $L$ and $\mu$ to vary. We allow $L$ to vary in the range of ${\mathcal O}(1)\times10^{-17} {\rm GeV^{-1}}$, $\mu$ in the range $0.8 - {\mathcal O}(1) \times 10^{17}$GeV, for the reasons mentioned above. It has turned out that there are values of $L$ and $\mu$ for which the data i) $-$ v) are reproduced precisely. The allowed values of $L$ and  $\mu$ are correlated. Their region is shown as a shaded area in the parameter space in fig.2. We take one point from this region,
\begin{equation}
       L=3.51 \times10^{-17} {\rm GeV^{-1}}, \ \mu=1 \times 10^{15} {\rm GeV},\label{ex} 
  \end{equation}
(the solid circle in fig.2), and show how the potential $V$ behaves as a function of $\phi$ in fig.3.
In fig.3, the region of inflation, $\phi_e\leq \phi \leq \phi_i$, is shown as a shaded area in which $\phi_i=1.60 \times 10^{19}  \mathrm{GeV}$ and $\phi_e=1.76 \times 10^{18} \mathrm{GeV}$.

The inflaton mass is mainly fixed by the circumference $L$, $m_{\phi}^2\simeq 1/(M_P^2L^4)$. For the parameter value of (\ref{ex}), $m_{\phi}=1.28 \times 10^{14}$GeV. The tensor-to-scalar ratio, evaluated at $\phi_i=1.60\times 10^{19}$GeV, $r=1.2\times 10^{-3}$ is consistent with the observations. 

{\bf II)} If we allow a large values of $c$, there are alternative solutions which reproduce the astrophysical data equally well. These solutions correspond to somewhat smaller values of $L$ and similar values of $\mu$. One example is : $c=51$ and 
\begin{equation}
      L=1\times10^{-16} {\rm GeV^{-1}}, \ \mu=1 \times 10^{15} {\rm GeV}.
\end{equation}

To conclude, we have found that the radion potential from one-loop corrections in 5D gravity with matter can give rise to inflation in accord with the astrophysical data. Gratifyingly, precise fine-tuning of the parameters is not necessary; there is an ample range of parameters $L$ and $\mu$ (shaded region in fig.2) for which desirable inflation should have occurred. 
\section{Discussion}
We have constructed an inflaton model by identifying the inflaton with the radion in 5D gravity with matter fermions. The model contains two parameters, the $S^1$ circumference $L$ of the fifth dimension and the fermion mass $\mu$, after choosing the number of fermions $c$ to be 2. There is an ample region in the parameter space for which the model can reproduce the astrophysical data. In this sense the radion model is free from extreme fine-tuning of the potential parameters, $L$ and $\mu$.

We have previously constructed an inflaton model based on 5D gauge theories. The inflaton is identified with the Higgs field, the fifth component of the gauge field. The model can reproduce the astrophysical data without fine-tuning of the parameters of the model. There remains a question about the value of the gauge coupling constant $g$; the value of $g$ has to be smaller than realistic values to explain the data. 

     In higher dimensional space-time both gauge field and gravity contain scalar fields, Higgs and radion respectively \cite{F}. It is interesting to note that the inflaton potential 
$V_{\rm grav}(\phi)$ arising from one-loop effects 
in the 5D gravity theory and $V_{\rm gauge}(\phi)$
arising from one-loop effects in the 5D gauge 
theory \cite{2,3} are very different. $V_{\rm gauge}(\phi)$
is periodic in $\phi$ whereas $V_{\rm gauge}(\phi)$ is 
not so, as seen in fig.1. This is because 
$V_{\rm gauge}(\phi)$ arises through the Wilson 
line. Presently we have no means to decide 
which kind of potential is more suitable as the 
inflaton potential. The two scalar fields may compete in giving rise to cosmological inflation. It is an interesting question which scalar field plays the dominant role. This question is currently studied by us.      
\ack
This work is supported partially by the grants for scientific research of the Ministry of Education, Kiban A, 21244036 and Kiban C, 20012487.

\end{document}